\documentclass{article}
\usepackage{amsmath,amssymb,amsfonts}

\def\op#1{\mathop{{\it\fam0} #1}\limits}

\newcommand{\pr}{{\mathrm pr}}

\newcommand{\Ker}{\mathrm{Ker}\,}

\newcommand{\beq}{\begin{equation}}
\newcommand{\eeq}{\end{equation}}
\newcommand{\ben}{\begin{eqnarray}}
\newcommand{\een}{\end{eqnarray}}
\newcommand{\be}{\begin{eqnarray*}}
\newcommand{\ee}{\end{eqnarray*}}
\newcommand{\bea}{\begin{eqalph}}
\newcommand{\eea}{\end{eqalph}}

\newcommand{\cB}{{\mathcal B}}

\newcommand{\cA}{{\mathcal A}}

\newcommand{\gE}{{\mathfrak E}}

\newcommand{\gT}{{\mathfrak T}}

\newcommand{\cL}{{\mathcal L}}

\newcommand{\cE}{{\mathcal E}}
\newcommand{\cH}{{\mathcal H}}

\newcommand{\cD}{{\mathcal D}}

\newcommand{\bL}{{\mathbf L}}
\newcommand{\bH}{{\mathbf H}}

\newcommand{\rrq}{{\ol q}}

\newcommand{\dl}{\delta}
\newcommand{\la}{\lambda}

\newcommand{\p}{\pi}

\newcommand{\Om}{\Omega}

\newcommand{\g}{\gamma}
\newcommand{\G}{\Gamma}

\newcommand{\Si}{\Sigma}

\newcommand{\w}{\wedge}

\newcommand{\wt}{\widetilde}
\newcommand{\wh}{\widehat}
\newcommand{\ol}{\overline}

\newcommand{\dr}{\partial}

\newcommand{\ar}{\op\longrightarrow}

\newcommand{\ot}{\otimes}
\newcommand{\ap}{\approx}

\newenvironment{eqalph}{\stepcounter{equation}
\setcounter{equationa}{\value{equation}} \setcounter{equation}{0}

\begin{eqnarray}}{\end{eqnarray}\setcounter{equation}{\value{equationa}}}

\begin{document}

\hbox{}

\begin{center}

{\Large\bf

Fibre bundle formulation of time-dependent mechanics}

\bigskip

G. SARDANASHVILY

\medskip

Department of Theoretical Physics, Moscow State University, 117234
Moscow, Russia

\bigskip

\end{center}

\begin{abstract}
We address classical and quantum mechanics in a general setting of
arbitrary time-dependent transformations. Classical
non-relativistic mechanics is formulated as a particular field
theory on smooth fibre bundles over a time axis $\mathbb R$.
Connections on these bundles describe reference frames. Quantum
time-dependent mechanics is phrased in geometric terms of Banach
and Hilbert bundles and connections on these bundles. A
quantization scheme speaking this language is geometric
quantization.
\end{abstract}

\section{Introduction}

The technique of symplectic manifolds is well known to provide the
adequate Hamiltonian formulation of autonomous mechanics
\cite{libe,vais}. Its realistic example is a mechanical system
whose configuration space is a manifold $M$ and whose phase space
is the cotangent bundle $T^*M$ of $M$ provided with the canonical
symplectic form $\Om_M= dp_i\w dq^i$, written with respect to the
holonomic coordinates $(q^i, p_i=\dot q_i)$ on $T^*M$. Any
autonomous Hamiltonian system locally is of this type.

However, this geometric formulation of autonomous mechanics is not
extended to mechanics under time-dependent transformations because
the symplectic form $\Om_M$ fails to be invariant under these
transformations. As a palliative variant, one has developed
time-dependent mechanics on a configuration space $Q=\mathbb
R\times M$ where $\mathbb R$ is the time axis \cite{eche,leon}.
Its phase space $\mathbb R\times T^*M$ is provided with the
pull-back presymplectic form $\pr^*_2\Om_M =dp_i\w dq^i$. However,
this presymplectic form also is broken by time-dependent
transformations.

We address non-relativistic mechanics in a case of arbitrary
time-dependent transformations \cite{arxiv09,book10,book98}. Its
configuration space is a fibre bundle $Q\to \mathbb R$ endowed
with bundle coordinates $(t,q^i)$, where $t$ is the standard
Cartesian coordinate on the time axis $\mathbb R$ with transition
functions $t'=t+$const. Its velocity space is the first order jet
manifold $J^1Q$ of sections of $Q\to \mathbb R$ coordinated by
$(t,q^i,q^i_t)$. A phase space is the vertical cotangent bundle
$V^*Q$ of $Q\to\mathbb R$ \cite{book10,sard98}.

This formulation of non-relativistic mechanics is similar to that
of classical field theory on fibre bundles over a base of
dimension $>1$ \cite{book09,sard08}. A difference between
mechanics and field theory however lies in the fact that
connections on bundles over $\mathbb R$ are flat, and they fail to
be dynamic variables, but describe reference frames.

Note that relativistic mechanics is adequately formulated as
particular classical string theory of one-dimensional submanifolds
\cite{book09,book10,rel1,rel2}.

In Section 6, time-dependent integrable Hamiltonian systems and
mechanics with time-dependent parameters are considered.

\section{Dynamic equations}

Let us start with the notion of a reference frame in
non-relativistic mechanics. A fibre bundle $Q\to\mathbb R$ always
is trivial. By the well known theorem \cite{book09,book00}, there
is one-to-one correspondence between the connections
\beq
\G =
\dr_t + \G^i \dr_i \label{a1.10}
\eeq
on $Q\to\mathbb R$ and the atlases of local constant
trivializations of $Q\to\mathbb R$ with time-independent
transition functions $q^i\to q'^i(q^j)$ so that $\G=\dr_t$ with
respect to an associated atlas. This fact leads to definition of a
reference frame in non-relativistic mechanics as a connection $\G$
on a configuration space $Q\to\mathbb R$
\cite{book10,massa,sard98}. The corresponding covariant
differential
\be
D^\G: J^1Q\ni \dr_t +q^i_t\dr_i\to (q^i_t-\G^i)\dr_i\in VQ
\ee
determines the relative velocity $(q^i_t-\G^i)\dr_i$ with respect
to a reference frame $\G$.

Equations of motion of non-relativistic mechanics usually are
first and second order dynamic equations
\cite{arxiv09,book10,book98}. A first order dynamic equation on a
fibre bundle $Q\to\mathbb R$ is a kernel of the covariant
differential $D^\G=(q^i_t-\G^i)\dr_i$ of some connection $\G$
(\ref{a1.10}) on $Q\to\mathbb R$. Second order dynamic equations
\beq
q^i_{tt}=\xi^i(t,q^j,q^j_t), \qquad \xi=\dr_t + q^i_t\dr_i + \xi^i
\dr_i^t, \label{z273}
\eeq
on $Q\to\mathbb R$ are conventionally defined as holonomic
connections $\xi$ on a jet bundle $J^1Q\to\mathbb R$. These
equations also are represented by connections
\be
\g=dq^\la\ot (\dr_\la + \g^i_\la \dr_i^t)
\ee
on an affine jet bundle $J^1Q\to Q$ and, due to the canonical
imbedding $J^1Q\to TQ$, they are equivalent to geodesic equations
on the tangent bundle $TQ$ of $Q$ \cite{book10,jmp00}.

One says that the second order dynamic equation (\ref{z273}) is a
free motion equation if there exists a reference frame $(t,\ol
q^i)$ on $Q$ such that this equation reads $\ol q^i_{tt}=0$.
Relative to an arbitrary frame $(t,q^i)$, a free motion equation
takes a form
\be
 q^i_{tt}=d_t\G^i +\dr_j\G^i(q^j_t-\G^j) -
\frac{\dr q^i}{\dr\rrq^m}\frac{\dr\rrq^m}{\dr q^j\dr
q^k}(q^j_t-\G^j) (q^k_t-\G^k),  \quad \G^i=\dr_t q^i(t,\ol q^j).
\ee
Its right-hand side is treated as an inertial force. One can show
that a free motion equation on a fibre bundle $Q\to\mathbb R$
exists iff $Q$ is a toroidal cylinder.

To consider a relative acceleration with respect to a reference
frame $\G$,  one should prolong a connection $\G$ on a
configuration space $Q\to\mathbb R$ to a holonomic connection
$\xi_\G$ on a jet bundle $J^1Q\to \mathbb R$. Given a second order
dynamic equation $\xi$, one can treat the vertical vector field
$a_\G\op=\xi-\xi_\G=(\xi^i-\xi_\G^i)\dr^t_i$ on $J^1Q\to Q$  as a
relative acceleration with respect to a frame $\G$. Then the
second order dynamic equation (\ref{z273}) can be written in a
covariant form $q^i_{tt} -\xi^i_\G=a_\G$ \cite{book10}.

\section{Lagrangian time-dependent mechanics}

Lagrangian mechanics is formulated in the framework of Lagrangian
formalism on fibre bundles \cite{book09,book10,sard13}. We
restrict our consideration to first order Lagrangian theory on a
fibre bundle $Q\to \mathbb R$ which is the case of
non-relativistic mechanics.

A first order Lagrangian is defined as a density
\beq
L=\cL dt, \qquad \cL: J^1Q\to \mathbb R, \label{23f2}
\eeq
on a velocity space $J^1Q$. There is the decomposition
\beq
dL=\dl L-d_H H_L, \label{+421}
\eeq
where we have the second-order Lagrange operator
\beq
\dl L= (\dr_i\cL- d_t\dr^t_i\cL) dq^i\w dt \label{305}
\eeq
and the Poincar\'e--Cartan form
\beq
H_L=\dr^t_i\cL dq^i -(q^i_t\dr^t_i\cL -\cL)dt. \label{303}
\eeq
A kernel of the Lagrange operator (\ref{305}) provides a second
order Lagrange equation
\beq
(\dr_i- d_t\dr^t_i)\cL=0. \label{b327}
\eeq

Every first order Lagrangian $L$ (\ref{23f2}) yields the Legendre
map
\beq
\wh L:J^1Q\ar_Q V^*Q,\qquad  p_i \circ\wh L = \pi_i=\dr^t_i\cL,
\label{a303}
\eeq
where $(t,q^i,p_i)$ are holonomic coordinates on the vertical
cotangent bundle $V^*Q$ of $Q\to\mathbb R$. A Lagrangian $L$ is
called hyperregular if $\wh L$ (\ref{a303}) is a diffeomorphism
and almost regular if a Lagrangian constraint space $N_L=\wh
L(J^1Q)$ is a closed imbedded subbundle of the Legendre bundle
$\pi_\Pi:V^*Q\to Q$ and the Legendre map $\wh L:J^1Q\to N_L$ is a
fibred manifold with connected fibres.

Besides the Lagrange equation (\ref{b327}), the Cartan equation
also is considered in Lagrangian mechanics. It is readily observed
that the Poincar\'e--Cartan form $H_L$ (\ref{303}) also is a
Poincar\'e--Cartan form of a first order Lagrangian
\be
\wt L=\wh h_0(H_L) = (\cL + (q_{(t)}^i - q_t^i)\p_i)dt, \qquad \wh
h_0(dq^i)=q^i_{(t)} dt,
\ee
on a repeated jet manifold $J^1J^1Y$ \cite{book09,book10}. The
Lagrange operator for $\wt L$ reads
\be
\dl\wt L  = [(\dr_i\cL - \wh d_t\p_i + \dr_i\p_j(q_{(t)}^j -
q_t^j))dq^i + \dr_i^t\p_j(q_{(t)}^j - q_t^j) dq_t^i]\w dt.
\ee
Its kernel $\Ker\dl \ol L\subset J^1J^1Q$ defines a first-order
Cartan equation
\beq
\dr_i^t\p_j(q_{(t)}^j - q_t^j)=0, \qquad \dr_i\cL - \wh d_t\p_i +
\dr_i\p_j(q_{(t)}^j - q_t^j)=0 \label{b336}
\eeq
on $J^1Q$. A key point is that the Cartan equation (\ref{b336}),
but not the Lagrange one (\ref{b327}) is associated to a Hamilton
equation in Hamiltonian mechanics.

The Poincar\'e--Cartan form $H_L$ (\ref{303}) yields a homogeneous
Legendre map $\wh H_L: J^1Q\to T^*Q$. Given holonomic coordinates
$(t,q^i,p_0,p_i)$ on $T^*Q$, it reads
\be
(p_0,p_i)\circ \wh H_L =(\cL-q^i_t\p_i, \p_i).
\ee
We have a one-dimensional affine bundle $\zeta:T^*Q\to V^*Q$ over
the vertical cotangent bundle $V^*Q$, and the Legendre map $\wh L$
(\ref{a303}) is the composition of morphisms $\wh L=\zeta\circ \wh
H_L.$ In comparison with a phase space $V^*Q$ of non-relativistic
mechanics, the cotangent bundle $T^*Q$ is its homogeneous phase
space.

In accordance with the first Noether theorem, Lagrangian
conservation laws in Lagrangian mechanics can be defined
\cite{book10,jmp07}. Let $u=u^t\dr_t +u^i\dr_i$, $u^t=0,1$, be a
vector field on a fibre bundle $Q\to\mathbb R$. The Lie derivative
$\bL_{J^1u} L$ of a Lagrangian $L$ along the jet prolongation
$J^1u$ of $u$ onto $J^1Q$ fulfils the first variational formula
\beq
\bL_{J^1u}L= u_V\rfloor\dl L + d_H(u\rfloor H_L), \label{23f42}
\eeq
which results from the decomposition (\ref{+421}). A vector field
$u$ is called a symmetry of a Lagrangian $L$ if the Lie derivative
$\bL_{J^1u} L$ vanishes. In  this case, the first variational
formula (\ref{23f42}) leads to a weak conservation law
\beq
0\ap d_t\gT_u, \qquad \gT_u=u\rfloor H_L=(u^i-u^tq^i_t)\pi_i +
u^t\cL, \label{gm488}
\eeq
of a symmetry current $\gT_u$ along a vector field $u$.

For instance, if $u^t=1$, we have a reference frame $u=\G$, and
the symmetry current (\ref{gm488}) is an energy function
\be
E_\G=-\gT_\G= \pi_i(q^i_t -\G^i) -\cL
\ee
relative to a reference frame $\G$ \cite{eche95,book10,sard98}.

\section{Hamiltonian time-dependent mechanics}

A phase space $V^*Q$ of Hamiltonian time-dependent mechanics is
provided with the canonical Poisson structure
\beq
\{f,g\}_V = \dr^if\dr_ig-\dr^ig\dr_if, \qquad f,g\in
C^\infty(V^*Q), \label{m72}
\eeq
such that $\zeta^*\{f,g\}_V=\{\zeta^*f,\zeta^*g\}_T$, where
$\{f,g\}_T$ is the Poisson bracket for the canonical symplectic
structure $\Om_Q$ on the cotangent bundle $T^*Q$ of $Q$.

However, Hamiltonian mechanics  is not familiar Poisson
Hamiltonian theory on a Poisson manifold $V^*Q$ because all
Hamiltonian vector fields on $V^*Q$ are vertical. Hamiltonian
mechanics on $V^*Q$ is formulated as particular (polysymplectic)
Hamiltonian formalism on fibre bundles \cite{jpa99,book09,book10}.
Its Hamiltonian is a global section
\beq
h:V^*Q\to T^*Q, \qquad p_0\circ h=\cH(t,q^j,p_j), \label{ws513}
\eeq
of an affine bundle $T^*Q\to V^*Q$. The pull-back $(-h)^*\Xi$ of
the canonical Liouville form $\Xi=p_\mu dq^\mu$ on $T^*Q$ with
respect to this section is a Hamiltonian one-form
\beq
H=(-h)^*\Xi= p_k dq^k -\cH dt  \label{b4210}
\eeq
on $V^*Q$ \cite{book10,sard98}. This is the well-known invariant
of Poincar\'e--Cartan \cite{arn}.

For instance, any connection $\G$ (\ref{a1.10}) on $Q\to\mathbb R$
defines the global section $h_\G=p_i\G^i$ (\ref{ws513}) of an
affine bundle $T^*Q\to V^*Q$ and the corresponding Hamiltonian
form
\beq
 H_\G= p_k dq^k -\cH_\G dt=
p_k dq^k -p_i\G^i dt. \label{ws515}
\eeq
Furthermore, given a connection $\G$, any Hamiltonian form
(\ref{b4210}) admits a splitting
\beq
H= H_\G -\cE_\G dt, \qquad \cE_\G=\cH-\cH_\G=\cH- p_i\G^i,
\label{m46'}
\eeq
where $\cE_\G$ is called the Hamiltonian function on $V^*Q$
relative to a frame $\G$.

Given the Hamiltonian form $H$ (\ref{b4210}), there exists a
unique connection
\be
\g_H=\dr_t + \dr^k\cH\dr_k- \dr_k\cH\dr^k,
\ee
on $V^*Q\to \mathbb R$ such that $\g_H\rfloor dH=0$. It yields a
first order Hamilton equation \index{Hamilton equation}
\beq
q^k_t=\dr^k\cH, \qquad  p_{tk}=-\dr_k\cH \label{z20}
\eeq
on $V^*Q\to\mathbb R$, where $(t,q^k,p_k,q^k_t,p_{tk})$ are the
adapted coordinates on $J^1V^*Q$.

Herewith, a time-dependent Hamiltonian system $(\cH, V^*Q)$ is
associated to the homogeneous autonomous Hamiltonian system with a
Hamiltonian $\cH^*=p_0+\cH$ on the cotangent bundle $T^*Q$ so that
the Hamilton equation (\ref{z20}) on $V^*Q$ is equivalent to an
autonomous Hamilton equation on $T^*Q$ \cite{dew,book10,mang00}.

Moreover, the Hamilton equation (\ref{z20}) on $V^*Q$ also is
equivalent to the Lagrange equation of a Lagrangian
\beq
L_H=h_0(H) = (p_iq^i_t - \cH)dt \label{Q33}
\eeq
on the jet manifold $J^1V^*Q$ of $V^*Q\to\mathbb R$
\cite{book10,jmp07,sard98}. As a consequence, Hamiltonian
conservation laws can be formulated as the Lagrangian ones. In
particular, any integral of motion $F$ of the Hamilton equation
(\ref{z20}) is a conserved current of the Lagrangian (\ref{Q33}),
and {\it vice versa}. It obeys the evolution equation
\beq
\bL_{\g_H} F=\dr_tF +\{\cH,F\}_V =0 \label{ws516}
\eeq
and, equivalently, the homogeneous evolution equation
\beq
\zeta^*(\bL_{\g_H}F)=\{\cH^*,\zeta^*F\}_T=0. \label{077}
\eeq

In particular, let $\cE_\G$ (\ref{m46'}) be a Hamiltonian function
relative to a reference frame $\G$. Given bundle coordinates
adapted to $\G$, its evolution equation (\ref{ws516}) takes a form
\be
\bL_{\g_H}\cE_\G=\dr_t\cE_\G=\dr_t\cH.
\ee
It follows that, a Hamiltonian function $\cE_\G$ relative to a
reference frame $\G$ is an integral of motion iff a Hamiltonian,
written with respect to $\G$, is time-independent. One can think
of $\cE_\G$ as being an energy function relative to a reference
frame $\G$ \cite{eche95,book10,jmp07,sard98}. Indeed, if $\cE_\G$
is an integral of motion, it is a conserved symmetry current of
the canonical lift onto $V^*Q$ of the vector field $-\G$
(\ref{a1.10}) on $Q$.

Lagrangian and Hamiltonian formulations of time-dependent
mechanics fail to be equivalent. The relations between Lagrangian
and Hamiltonian formalisms are based on the facts that: (i) every
first order Lagrangian $L$ (\ref{23f2}) on a velocity space $J^1Q$
induces the Legendre map (\ref{a303}) of this velocity space to a
phase space $V^*Q$, (ii) every Hamiltonian form $H$ (\ref{b4210})
on a phase space $V^*Q$ yields a Hamiltonian map
\be
\wh H: V^*Q\ar J^1Q, \qquad q^i_t\circ\wh H=\dr^i\cH,
\ee
of this phase space to a velocity space $J^1Q$.

Given a Lagrangian $L$, the Hamiltonian form $H$ (\ref{b4210}) is
said to be associated with $L$ if $H$ satisfies the relations
\beq
\wh L\circ\wh H\circ \wh L=\wh L,\qquad \wh H^*L_H=\wh H^*L,
\label{d2.30}
\eeq
where $L_H$ is the Lagrangian (\ref{Q33}).

For instance, let $L$ be a hyperregular Lagrangian. It follows
from the relations (\ref{d2.30}) that, in this case, $\wh H=\wh
L^{-1}$ and there exists a unique Hamiltonian form
\beq
H=p_kdq^k-\cH dt, \qquad \cH=p_i\wh L^{-1i} - \cL(t, q^j,\wh
L^{-1j}), \label{cc311}
\eeq
associated with $L$. Let $s$ be a solution of the Lagrange
equation (\ref{b327}) for a Lagrangian $L$. A direct computation
shows that $\wh L\circ J^1s$ is a solution of the Hamilton
equation (\ref{z20}) for the Hamiltonian form $H$ (\ref{cc311}).
Conversely, if $r$ is a solution of the Hamilton equation
(\ref{z20}) for the Hamiltonian form $H$ (\ref{cc311}), then
$s=\pi_\Pi\circ r$ is a solution of the Lagrange equation
(\ref{b327}) for $L$. It follows that, in the case of hyperregular
Lagrangians, Hamiltonian formalism is equivalent to Lagrangian
one.

If a Lagrangian is not hyperregular, an associated Hamiltonian
form need not exist or it is not unique. Comprehensive relations
between Lagrangian and Hamiltonian systems are established in the
case of almost regular Lagrangians \cite{book10,mang00,sard98}.

\section{Quantum time-dependent mechanics}

Quantum time-dependent mechanics is phrased in geometric terms of
Banach and Hilbert manifolds and Hilbert and $C^*$-algebra
bundles. Quantization schemes speaking this language are
instantwise and geometric quantizations
\cite{jmp02a,book05,book10}.

A definition of smooth Banach and Hilbert manifolds follows that
of the finite-dimensional ones, but Banach manifolds are not
locally compact, and they need not be paracompact
\cite{book05,lang95,vais73}. It is essential that Hilbert
manifolds satisfy the inverse function theorem and, therefore,
locally trivial Hilbert bundles are defined. However, the
following fact leads to the non-equivalence of Schr\"odinger and
Heisenberg quantization. Let $E$ a Hilbert space and $B$ some
$C^*$-algebra of bounded operators in $E$. There is a topological
obstruction to the existence of associated Hilbert and
$C^*$-algebra bundles $\cE$ and $\cB$ with typical fibres $E$ and
$B$, respectively. Firstly, transition functions of $\cE$ define
those of $\cB$, but the latter are not continuous in general.
Secondly, transition functions of $\cB$ need not give rise to
those of $\cE$.

One also meets a problem of the definition of connections on
$C^*$-algebra bundles. It comes from the fact that a $C^*$-algebra
need not admit non-zero bounded derivations. An unbounded
derivation  of a $C^*$-algebra $A$ obeying certain conditions is
an infinitesimal generator of a strongly (but not uniformly)
continuous one-parameter group of automorphisms of $A$
\cite{brat75,book05,book10}. Therefore, one must introduce a
connection on a $C^*$-algebra bundle in terms of parallel
transport operators, but not their infinitesimal generators
\cite{asor,book05}. Moreover, a representation of $A$ need not
imply a unitary representation of its strongly continuous
one-parameter group of automorphisms. In contrast, connections on
a Hilbert bundle over a smooth manifold can be defined as first
order differential operators on a module of its sections
\cite{book05,book10}.

In particular, this is the case of instantwise quantization
describing evolution of quantum systems in terms of Hilbert
bundles over $\mathbb R$ \cite{jmp02,book05,book10,sard00}.
Namely, let us consider a Hilbert bundle $\gE\to \mathbb R$ with a
typical fibre $E$ and a connection $\nabla_t$ on a
$C^\infty(\mathbb R)$-module $\gE(\mathbb R)$ of smooth sections
of $\gE\to\mathbb R$. It obeys the Leibniz rule
\be
\nabla_t (f\psi)= \dr_tf\psi+ f\nabla_t \psi, \qquad \psi\in
\gE(\mathbb R), \qquad f\in C^\infty(\mathbb R).
\ee
Given a trivialization $\gE=\mathbb R\times E$, the connection
$\nabla_t$ reads
\beq
\nabla_t\psi =(\dr_t + i\bH(t)) \psi, \label{+346}
\eeq
where $\bH(t)$ are bounded self-adjoint operators in $E$ for all
$t\in\mathbb R$. A section $\psi$ of $\gE\to\mathbb R$ is an
integral section of the connection $\nabla_t$ (\ref{+346}) if it
obeys the equation
\beq
\nabla_t\psi(t) =(\dr_t + i\bH(t)) \psi(t)=0. \label{+349}
\eeq
One can think of this equation as being the Schr\"odinger
equation.

The most of quantum models come from canonical quantization of
classical mechanical systems by means of replacement of a Poisson
bracket $\{f,f'\}$ of smooth functions with a bracket $[\wh f,\wh
f']$ of Hermitian operators in a Hilbert space in accordance with
Dirac's condition $[\wh f,\wh f']=-i\wh{\{f,f'\}}$. Canonical
quantization of Hamiltonian time-dependent mechanics on a
configuration space $Q\to\mathbb R$ is geometric quantization
\cite{jmp02a,jmp02,book05,book10}. A key point is that, in this
case, the evolution equation (\ref{ws516}) is not reduced to the
Poisson bracket on a phase space $V^*Q$, but is expressed as
(\ref{077}) in the Poisson bracket on the homogeneous phase space
$T^*Q$. Therefore, the compatible geometric quantization both of
the symplectic cotangent bundle $T^*Q$ and the Poisson vertical
cotangent bundle $V^*Q$ of $Q$ is required.

Note that geometric quantization of Poisson manifolds is
formulated in terms of contravariant connections \cite{vais}.
Though there is one-to-one correspondence between the Poisson
structures on a manifold and its symplectic foliations, this
quantization of a Poisson manifold need not imply quantization of
its symplectic leaves \cite{vais97}. Geometric quantization of
symplectic foliations disposes of this problem
\cite{jmp02,book05,book10,sard01}. A quantum algebra of a
symplectic foliation also is that of an associated Poisson
manifold whose restriction to each symplectic leaf is its quantum
algebra.

Namely, the standard prequantization of the cotangent bundle
$T^*Q$ yields the compatible prequantization of a Poisson manifold
$V^*Q$. However, polarization of $T^*Q$ need not induce any
polarization of $V^*Q$, unless it contains the vertical cotangent
bundle of a fibre bundle $T^*Q\to V^*Q$ spanned by vectors
$\dr^0$. A unique canonical real polarization of $T^*Q$,
satisfying this condition, is the vertical tangent bundle of
$T^*Q\to Q$. The associated quantum algebra $\cA_T$ consists of
functions on $T^*Q$ which are affine in momenta $p_\mu$. This
polarization of $T^*Q$ yields polarization of a Poisson manifold
$V^*Q$ such that the corresponding quantum algebra $\cA_V$
consists of functions on $V^*Q$ which are affine in momenta $p_i$,
i.e., $\cA_V$ is a subalgebra of $\cA_T$. After metaplectic
correction, we obtain compatible Schr\"odinger representations
\ben
&& \wh f\rho=\left(-ia^\la\dr_\la
-\frac{i}{2}\dr_\la a^\la- b\right)
\rho,\qquad f=a^\la(q^\mu)p_\la + b(q^\mu)\in \cA_T, \label{qq82'}\\
&& \wh f\rho=\left(-ia^k\dr_k
-\frac{i}{2}\dr_k a^k- b\right) \rho,\qquad f=a^k(q^\mu)p_k +
b(q^\mu)\in \cA_V, \label{qq83}
\een
of $\cA_T$  and $\cA_V$ in the space $\cD_{1/2}(Q)$ of complex
half-densities $\rho$ on $Q$.

The Schr\"odinger quantization (\ref{qq83}) of $V^*Q$ provides
instantwise quantization of time-dependent mechanics
\cite{book10}. Indeed, a glance at the Poisson bracket (\ref{m72})
shows that the Poisson algebra $C^\infty(V^*Q)$ is a Lie algebra
over the ring $C^\infty(\mathbb R)$ of functions of time, where
algebraic operations in fact are instantwise operations depending
on time as a parameter. One can show that the Schr\"odinger
quantization (\ref{qq83}) of a Poisson manifold $V^*Q$ yields
geometric quantization of its symplectic fibres $V^*_tQ$,
$t\in\mathbb R$, such that the quantum algebra $\cA_t$ of $V_t^*Q$
consists of elements $f\in\cA_V$ restricted to $V_t^*Q$. Bearing
in mind that $\rho\in \cD_{1/2}[Q]$ are fibrewise half-densities
on $Q\to\mathbb R$, let us choose a carrier space of the
Schr\"odinger representation (\ref{qq83}) of $\cA_V$ which
consists of complex half-densities $\rho$ on $Q$ such that $\rho$
on $Q_t$ for any $t\in \mathbb R$ is of compact support. It is a
pre-Hilbert $C^\infty(\mathbb R)$-module $\gE_R$ which also is a
carrier space for the quantum algebra $\cA_T$, but its action in
$\gE_R$ is not instantwise.

Let us turn to quantization of an evolution equation. Since the
equation (\ref{ws516}) is not reduced to a Poisson bracket,
quantization of a Poisson manifold $V^*Q$ fails to provide
quantization of this evolution equation. Therefore, we quantize
the equivalent homogeneous evolution equation (\ref{077}) on a
symplectic manifold $T^*Q$. The Schr\"odinger representation
(\ref{qq82'}) of a Lie algebra $\cA_T$ is extended to its
enveloping algebra, and defines the quantization $\wh\cH^*$ of a
homogeneous Hamiltonian $\cH^*$. Moreover, since $\wh
p_0=-i\dr_t$, an operator $i\wh\cH^*$ obeys the Leibniz rule
\be
i\wh\cH^*(r\rho)=\dr_tr\rho +r(i\wh\cH^*\rho), \qquad r\in
C^\infty(\mathbb R), \qquad \rho\in \gE_R.
\ee
Thus, it is a connection on a $C^\infty(\mathbb R)$-module
$\gE_R$. Then a quantum constraint
\beq
i\wh\cH^*\rho=0, \qquad \rho\in \gE_R, \label{gg}
\eeq
is the Schr\"odinger equation (\ref{+349}) in quantum
time-dependent mechanics.

This quantization depends on a reference frame as follows. In
accordance with the Schr\"odinger representation (\ref{qq82'}), a
homogeneous Hamiltonian $\cH^*=p_0 +\cH$ is quantized as a
Hamilton operator
\beq
\wh\cH^*=\wh p_0+\wh\cH=-i\dr_t +\wh\cH. \label{j2}
\eeq
A problem is that the decomposition $\cH^*=p_0 +\cH$ and the
corresponding splitting (\ref{j2}) of a Hamilton operator
$\wh\cH^*$ are ill defined. At the same time, any reference frame
$\G$ yields the decomposition
\be
\cH^*=(p_0+\cH_\G) + (\cH-\cH_\G) = \cH^*_\G +\cE_\G,
\ee
where $\cH_\G$ is the Hamiltonian (\ref{ws515}) and $\cE_\G$
(\ref{m46'}) is an energy function relative to a reference frame
$\G$. Accordingly, we obtain the splitting of a Hamilton operator
\be
\wh\cH^*=\wh\cH_\G^* +\wh\cE_\G, \qquad \wh\cH^*_\G =-i\dr_t
-i\G^k\dr_k -\frac{i}{2}\dr_k\G^k
\ee
and $\wh\cE_\G$ is the operator of energy relative to a reference
frame $\G$ \cite{book10,jmp07}. Given a reference frame $\G$, the
energy function $\cE_\G$ is quantized as $\wh \cE_\G=\wh\cH^*-
\wh\cH^*_\G$.  As a consequence, the Schr\"odinger equation
(\ref{gg}) reads
\be
(\wh\cH_\G +\wh\cE_\G)\rho=-i\left(\dr_t +\G^k\dr_k
+\frac12\dr_k\G^k\right)\rho +\wh\cE_\G\rho=0.
\ee

\section{Outcomes}

The Liouville--Arnold theorem for completely integrable systems
and the Mishchenko--Fomenko theorem for the superintegrable ones
state the existence of action-angle coordinates around a compact
invariant submanifold of a Hamiltonian integrable system. These
theorems have been generalized to the case of non-compact
invariant submanifolds
\cite{fior,fior2,jmp07a,jmp03,book05,book10,ijgmmp09a}. In
particular, this is the case of time-dependent completely
integrable and superintegrable systems
\cite{acang2,book05,book10,sard12}. Geometric quantization of
completely integrable and superintegrable Hamiltonian systems with
respect to action-angle variables has been considered
\cite{acang1,plet02,book05,pl07,book10}.

At present, quantum systems with classical parameters attract
special attention in connection with holonomic quantum
computation. These parameters can be seen as sections of some
smooth fibre bundle $\Si\to\mathbb R$. Then a configuration space
of a mechanical system with time-dependent parameters is a
composite fibre bundle $Q\to\Si\to\mathbb R$
\cite{jmp02,book05,book10,sard00}. The corresponding total
velocity and phase spaces are the first order jet manifold $J^1Q$
and the vertical cotangent bundle $V^*Q$ of the configuration
bundle $Q\to\mathbb R$, respectively. However, since parameters
are classical, a phase space of a quantum system with
time-dependent parameters is the vertical cotangent bundle
$V_\Si^*Q$ of a fibre bundle $Q\to\Si$. We apply to
$V_\Si^*Q\to\Si$ the technique of leafwise geometric quantization
\cite{jmp02,book05,book10}.

Geometric Berry's phase factor is a phenomenon peculiar to quantum
systems with classical parameters. It is characterized by a
holonomy operator driving a carrier Hilbert space over a parameter
manifold. A problem lies in separation of a geometric phase factor
from an evolution operator without using an adiabatic assumption.
Therefore, we address the Berry phase phenomena in completely
integrable systems. A reason is that, being constant under an
internal dynamic evolution, action variables of a completely
integrable system are driven only by a perturbation holonomy
operator without any adiabatic approximation
\cite{jmp04,book05,book10}.


\begin{thebibliography}{ederf}


\bibitem{arn} V.Arnold (Ed.), \emph{Dynamical Systems III, IV}
(Springer, Berlin, 1990).

\bibitem{asor} M.Asorey, J.Cari\~nena and M.Paramion, Quantum evolution as a
parallel transport, \emph{J. Math. Phys.} \textbf{23} (1982) 1451.

\bibitem{brat75} O.Bratteli and D.Robinson, Unbounded derivations of
$C^*$-algebras, \emph{Commun. Math. Phys.} \textbf{42} (1975) 253.

\bibitem{dew} A.Dewisme and S.Bouquet, First integrals and
symmetries of time-dependent Hamiltonian systems, \emph{J. Math.
Phys} \textbf{34} (1993) 997.


\bibitem{eche} A.Echeverr\'{\i}a Enr\'{\i}quez, M.Mu\~noz Lecanda and
N. Rom\'an Roy, Geometrical setting of time-dependent regular
systems. Alternative models, \emph{Rev. Math. Phys.} \textbf{3}
(1991) 301.

\bibitem{eche95} A.Echeverr\'{\i}a Enr\'{\i}quez, M.Mu\~noz Lecanda and
N. Rom\'an Roy,  Non-standard connections in classical mechanics,
\emph{J. Phys. A} \textbf{28} (1995) 5553.

\bibitem{acang1} E. Fiorani, G. Giachetta and G. Sardanashvily,
 Geometric quantization of time-dependent completely integrable
Hamiltonian systems,  \emph{J. Math. Phys.} \textbf{43} (2002)
5013; \emph{arXiv}: quant-ph/0202093.


\bibitem{fior} E.Fiorani, G.Giachetta, and G.Sardanashvily, The
Liouville -- Arnold -- Nekhoroshev theorem for noncompact
invariant manifolds, \emph{J. Phys. A} \textbf{36} (2003) L101;
\emph{arXiv}: math.DS/0210346.

\bibitem{fior2} E.Fiorani and G.Sardanashvily, Noncommutative integrability on
noncompact invariant manifold, \emph{J. Phys. A} \textbf{39}
(2006) 14035; \emph{arXiv}: math.DS/0604104.

\bibitem{jmp07a} E.Fiorani and G.Sardanashvily,
 Global action-angle coordinates for completely integrable systems
with noncompact invariant manifolds, \emph{J. Math. Phys.}
\textbf{48} (2007) 032001; \emph{arXiv}: math.DS/0610790.


\bibitem{jpa99} G.Giachetta, L.Mangiarotti and G.Sardanashvily, Covariant
Hamilton equations for field theory, \emph{J. Phys. A} \textbf{32}
(1999) 6629.

\bibitem{jmp02a} G.Giachetta, L.Mangiarotti and G.Sardanashvily,
Covariant geometric quantization of nonrelativistic time-dependent
mechanics, \emph{J. Math. Phys} \textbf{43} (2002) 56;
\emph{arXiv}: quant-ph/0012036.

\bibitem{jmp02} G.Giachetta, L.Mangiarotti and G.Sardanashvily,
Geometric quantization of mechanical systems with time-dependent
parameters, \emph{J. Math. Phys} \textbf{43} (2002) 2882;
\emph{arXiv}: quant-ph/0112011.

\bibitem{acang2} G.Giachetta, L.Mangiarotti and G.Sardanashvily,
Action-angle coordinates for time-dependent completely integrable
Hamiltonian systems, \emph{J. Phys. A} \textbf{35} (2002) L439;
\emph{arXiv}: math.DS/0204151.

\bibitem{plet02} G.Giachetta, L.Mangiarotti and G.Sardanashvily,
Geometric quantization of completely integrable systems in
action-angle variables, \emph{Phys. Lett. A} \textbf{301} (2002)
53; \emph{arXiv}: quant-ph/0112083.

\bibitem{jmp03} G.Giachetta, L.Mangiarotti and G.Sardanashvily,
 Bi-Hamiltonian partially integrable systems, \emph{J. Math.
Phys.} \textbf{44} (2003) 1984; \emph{arXiv}: math.DS/0211463.

\bibitem{jmp04}  G.Giachetta, L.Mangiarotti and
G.Sardanashvily, Nonadiabatic holonomy operators in classical and
quantum completely integrable systems, \emph{J. Math. Phys}
\textbf{45} (2004) 76; \emph{arXiv}: quant-ph/0212108.


\bibitem{book05} G.Giachetta, L.Mangiarotti and G.Sardanashvily,
\emph{Geometric and Algebraic Topological Methods in Quantum
Mechanics} (World Scientific, Singapore, 2005).

\bibitem{pl07} G.Giachetta, L.Mangiarotti and G.Sardanashvily,
 Quantization of noncommutative completely integrable Hamiltonian
systems, \emph{Phys. Lett. A} \textbf{362} (2007) 138;
\emph{arXiv}: quant-ph/0604151.

\bibitem{arxiv09} G.Giachetta, L.Mangiarotti, G.Sardanashvily, Advanced mechanics. Mathematical
introduction, \emph{arXiv}: 0911.0411.

\bibitem{book09} G.Giachetta, L.Mangiarotti and G.Sardanashvily,
\emph{Advanced Classical Field Theory} (World Scientific,
Singapore, 2009).

\bibitem{book10} G.Giachetta, L.Mangiarotti and G.Sardanashvily,
\emph{Geometric Formulation of Classical and Quantum Mechanics}
(World Scientific, Singapore, 2010).

\bibitem{leon} M.De Le\'on and P.Rodrigues \emph{Methods of Differential
Geometry in Analytical Mechanics} (North-Holland, Amsterdam,
1989).

\bibitem{lang95} S.Lang, \emph{Differential and Riemannian
Manifolds}, Gradutate Texts in Mathematics \textbf{160} (Springer,
New York, 1995).


\bibitem{libe} P.Libermann and C-M.Marle \emph{Symplectic Geometry and
Analytical Mechanics} (D.Reidel Publishing Company, Dordrecht,
1987).

\bibitem{book98} L.Mangiarotti and G.Sardanashvily \emph{Gauge
Mechanics} (World Scientific, Singapore, 1998).

\bibitem{book00} L.Mangiarotti and G.Sardanashvily, \emph{Connections in
Classical and Quantum Field Theory} (World Scientific, Singapore,
2000).

\bibitem{jmp00} L.Mangiarotti and G.Sardanashvily,
On the geodesic form of second order dynamic equations, \emph{J.
Math. Phys.} \textbf{41} (2000) 835.

\bibitem{mang00} L.Mangiarotti and G.Sardanashvily, Constraints
in Hamiltonian time-dependent mechanics, \emph{J. Math. Phys.}
\textbf{41} (2000) 2858; \emph{arXiv}: math-ph/9904028.


\bibitem{jmp07} L.Mangiarotti and G.Sardanashvily, Quantum mechanics with respect to different reference
frames, \emph{J. Math. Phys.} \textbf{48} (2007) 082104;
\emph{arXiv}: quant-ph/0703266.

\bibitem{massa} E.Massa and E.Pagani, Jet bundle geometry, dynamical
connections and the inverse problem of Lagrangian mechanics,
\emph{Ann. Inst. Henri Poincar\'e} \textbf{61} (1994) 17.

\bibitem{sard98} G.Sardanashvily, Hamiltonian time-dependent mechanics,
\emph{J. Math. Phys.} \textbf{39} (1998) 2714.

\bibitem{sard00} G.Sardanashvily, Classical and quantum mechanics with
time-dependent parameters, \emph{J. Math. Phys.} \textbf{41}
(2000) 5245.

\bibitem{sard01} G.Sardanashvily, Geometric quantization of symplectic
foliations, \emph{arXiv}: math.DG/0110196.

\bibitem{sard08} G.Sardanashvily, Classical field theory.
Advanced mathematical formulation, \emph{Int. J. Geom. Methods
Mod. Phys.} \textbf{5} (2008) 1163; \emph{arXiv}: 0811.0331.

\bibitem{ijgmmp09a}  G.Sardanashvily, Superintegrable Hamiltonian systems with
noncompact invariant submanifolds. Kepler system, \emph{Int. J.
Geom. Methods Mod. Phys.} \textbf{6} (2009) 1391; \emph{arXiv}:
0911.0992.

\bibitem{rel1} G.Sardanashvily, Relativistic mechanics in a general
setting, \emph{Int. J. Geom. Methods Mod. Phys.} \textbf{7} (2010)
1307; \emph{arXiv}: 1005.1212.

\bibitem{rel2} G.Sardanashvily, Lagrangian dynamics of submanifolds. Relativistic mechanics,
 \emph{J. Geom. Mech.} \textbf{4} (2012) 99; \emph{arXiv}: 1112.0216.

\bibitem{sard12} G.Sardanashvily, Time-dependent superintegrable Hamiltonian systems,
\emph{Int. J. Geom. Methods Mod. Phys.} \textbf{9} (2012) N8
1220016.

\bibitem{sard13} G.Sardanashvily, Graded Lagrangian formalism, \emph{Int. J. Geom. Methods Mod. Phys.} \textbf{10} (2013)
N5 1350016.

\bibitem{vais73} I.Vaisman, \emph{Cohomology and Differential Forms}
(Marcel Dekker, Inc., New York, 1973).

\bibitem{vais} I.Vaisman \emph{Lectures on the Geometry of Poisson Manifolds}
(Birkh\"auser, Basel, 1994).

\bibitem{vais97} I.Vaisman, On the geometric quantization of
the symplectic leaves of Poisson manifolds, \emph{Diff. Geom.
Appl.} \textbf{7} (1997) 265.





\end{thebibliography}
\end{document}